\documentclass[%
superscriptaddress,
bibnotes,
 amsmath,amssymb,
 aps,
 prl, 
eprint,
floatfix,
reprint
]{revtex4-2}
\usepackage{graphicx,xcolor}
\definecolor{darkblue}{RGB}{0,0,150}
\definecolor{nightblue}{RGB}{0,0,100}

\usepackage{mathrsfs,dsfont,mathtools}

\newcommand{\refsub}[2]{\hyperref[#1]{\ref*{#1}#2}}

\usepackage{dcolumn}
\usepackage{bm}
\usepackage[
colorlinks,
citecolor=darkblue,
linkcolor=darkblue,
urlcolor=nightblue]{hyperref}

\usepackage[english]{babel}
\usepackage[babel,kerning=true,spacing=true]{microtype}

\renewcommand{\Re}{\mathrm{Re}\,}
\renewcommand{\Im}{\mathrm{Im}\,}

\definecolor{DarkRed}{RGB}{100,0,0}
\definecolor{LightGreen}{RGB}{000,50,0}

\begin{document}

\title{
Para-hydrodynamics from weak surface scattering in 
ultraclean thin flakes
}

\author{Yotam Wolf} 
\author{Amit Aharon-Steinberg} 
\author{Binghai Yan}
\author{Tobias Holder}
\email{tobias.holder@weizmann.ac.il}
\affiliation{Department of Condensed Matter Physics,
Weizmann Institute of Science, Rehovot, Israel}

\date{\today}

\begin{abstract}
Electron hydrodynamics typically emerges in electron fluids with a high electron-electron collision rate. However, new experiments with thin flakes of WTe$_2$ have revealed that other momentum-conserving scattering processes can replace the role of the electron-electron interaction, thereby leading to a novel, so-called para-hydrodynamic regime. Here, we develop the kinetic theory for para-hydrodynamic transport. To this end, we consider a ballistic electron gas in a thin 3-dimensional sheet where the momentum-relaxing ($\ell_{mr}$) and momentum-conserving ($\ell_{mc}$) mean free paths are decreased due to boundary scattering from a rough surface.
The resulting effective mean free path of the in-plane components of the electronic flow are then expressed in terms of microscopic parameters of the sheet boundaries, predicting that a para-hydrodynamic regime with $\ell_{mr}\gg \ell_{mc}$ emerges generically in ultraclean three-dimensional materials. Using our approach, we recover the transport properties of WTe$_2$ in the para-hydrodynamic regime in good agreement with existing experiments.

\end{abstract}

\maketitle
\section{Introduction}
The viscous flow of an interacting electron fluid has been predicted a  long time ago~\cite{Gurzhi1968}, however, experimental evidence for it has remained scarce for many decades~\cite{deJong1995}.
The advent of ultra clean quantum materials with low carrier density~\cite{Narozhny2019,Yan2017,Armitage2018}, 
yielded a growing number of cases demonstrating viscous electron flow in the last few years ~\cite{KrishnaKumar2017,Moll2016,Gooth2017,Berdyugin2019,Osterhoudt2021,Kumar2021a}. Moreover, it has even become possible to establish ballistic and viscous flow profiles using spatially resolved techniques~\cite{Bandurin2016,Marguerite2019,Sulpizio2019,Ku2020,Vool2021}.
Among the coveted properties of hydrodynamic flow are for example a negative nonlocal resistance~\cite{Levin2018,Bandurin2018,Kim2020b,Gupta2021} 
as well as other nonlocal transport signatures~\cite{Guo2017,Alekseev2016,Torre2015,Mendoza2011,Stern2021,DiSante2020,Raichev2020}.
However, the observation of vortical flow (electron whirlpools) had still remained elusive~\cite{Falkovich2017,Nazaryan2021,Pellegrino2016,Alekseev2018a}.
This situation was upended very recently when a high-fidelity, spatially-resolved experiment~\cite{AharonSteinberg2022} in ultraclean WTe$_2$~\cite{Wang2015a,Woods2016} could unambiguously demonstrate hydrodynamic vortical flow,  clearly excluding a ballistic origin of the observed whirlpool pattern.

The observation of this hydrodynamical vortical flow is remarkable for two additional reasons. Firstly, the device is not an effective 2d-system, but a thin (thickness $d=48\mathrm{nm}$, width $w=550\mathrm{nm}$), three-dimensional flake exfoliated and fabricated from WTe$_2$ flakes with a very large bulk mean free path $\ell\approx 20\mathrm{\mu m}$. 
Secondly, at the measured temperature $T=4.5\mathrm{K}$ the electron-electron interaction leads to a momentum-conserving mean free path $\ell_{ee}\approx 10\mathrm{\mu m}\gg w$.
This means that the sample cannot be in the hydrodynamic regime, which is characterized by the condition
that $\ell_{ee}\ll w\ll \ell$~\cite{Gurzhi1968}.

\begin{figure}[h!]
	\centering
	\includegraphics[width=\columnwidth]{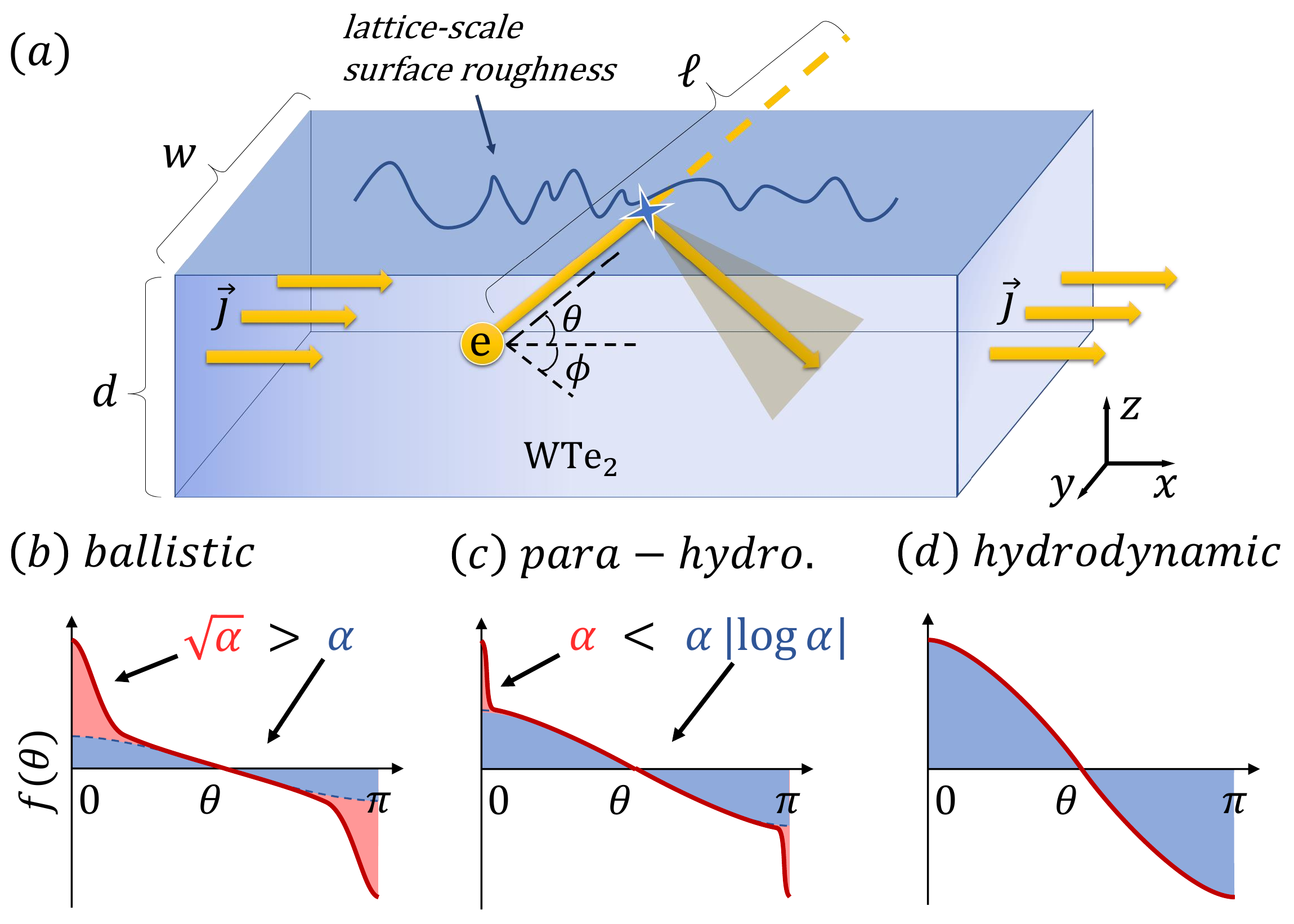}
	\caption{Phenomenology of para-hydrodynamics. 
		(a) The current flows through a thin slab of thickness $d\ll w\ll \ell$. Electrons scatter at the microscopically rough top and bottom surfaces with incident angle $\theta$. Most trajectories are reflected almost specularly, leading to angular diffusion in $\theta$, while few trajectories scatter randomly, thus dissipating momentum.
		(b-d) Comparison of the non-equilibrium distribution functions $f(\theta)$ in different flow regimes (red curve). An approximate cosine which inscribes $f(\theta)$ for steep angles $\theta\neq 0,\pi$ is shown as a blue shaded area, whereas the red shaded area indicates grazing trajectories ($\theta\approx 0,\pi$). 
		Different parts of $f(\theta)$ have dissimilar scale dependencies in terms of the ratio $\alpha=d/2\ell\ll 1$.
		For ballistic flow, $f(\theta)$ decays strongly upon approaching $\theta=\pi/2$, while for hydrodynamic flow it assumes a smooth cosine-form everywhere. The intermediate para-hydrodynamic regime carries signatures of both ballistic and hydrodynamic flow, but the  smooth part of $f(\theta)$ is logarithmically larger than the ballistic part, thus leading to an effectively hydrodynamic current in the limit $\alpha\to 0$.
	}
	\label{fig:schematics}
\end{figure}

Based on the unusual properties of WTe$_2$, Ref.~\cite{AharonSteinberg2022} suggested that an effectively hydrodynamic flow could instead be induced by almost specular (and thus predominantly momentum-conserving) scattering from the top and bottom surfaces of the three-dimensional WTe$_2$ sheet, a mechanism termed para-hydrodynamics (cf. Fig.~\refsub{fig:schematics}{a}). 
However, no kinetic theory has been proposed how this novel type of hydrodynamic flow emerges microscopically.

In this letter, we consider a kinetic theory for the in-plane flow of an electron fluid in a thin, three-dimensional slab that takes into account weak boundary scattering from rough top and bottom surfaces. We demonstrate that this setting naturally leads to para-hydrodynamic flow in thin three-dimensional devices as long as the bulk mean free path $\ell$ of the material is very large compared to the device thickness $d$.
The proposed scattering model is generic, with the only parameters being the amplitude and correlation length of the surface roughness.

The mechanism is depicted schematically in Fig.~\refsub{fig:schematics}{b-d}. The  emergence of the para-hydrodynamic regime is a result of the conversion of rarely colliding trajectories which impact the top and bottom surfaces at a grazing angle (i.e. ballistic flow) into trajectories that scatter often and at a steep angle from the surface (hydrodynamic flow). This conversion happens due to a slow angular diffusion of the scattered trajectories. 
The redistribution of statistical weight means that the distribution function and thus also the current predominantly resembles hydrodynamic transport.
Therefore, while it is not possible to describe the three-dimensional distribution function using a Stokes-Ohm hydrodynamic approach,  the in-plane components of the flow velocity exhibit relaxation properties which resemble viscous flow. Specifically,
we find that the effective in-plane momentum-relaxing ($\ell_{mr}$) and momentum-conserving ($\ell_{mc}$) mean free paths  are related as $\ell_{mr}/\ell_{mc}\propto\log(\ell/d)>1$.
Since the conversion of forward trajectories into steep trajectories is bound to happen whenever angular diffusion is present in the boundary scattering, the only reason why the phenomenon of para-hydrodynamics has so far remained elusive seems to be the logarithmically slow enhancement of this effect with increasing mean free path.
In particular, our model posits that neither the precise surface roughness nor the material itself sensitively affect whether a para-hydrodynamic transport regime can emerge in a given material. Instead, the para-hydrodynamic regime merely requires a very large fineness ratio $\ell/d$, combined with an appropriate choice of the width $w$ of the slab so that $\ell_{mr}\gg w \gg \ell_{mc}$. These conditions were fulfilled in the experiment of Ref.~\cite{AharonSteinberg2022}, which was done with high-quality samples where $\ell/d> 500$.

\section{Results}
\subsection{Absence of e-e interactions at low T}
\begin{figure}
	\centering
	\includegraphics[width=\columnwidth]{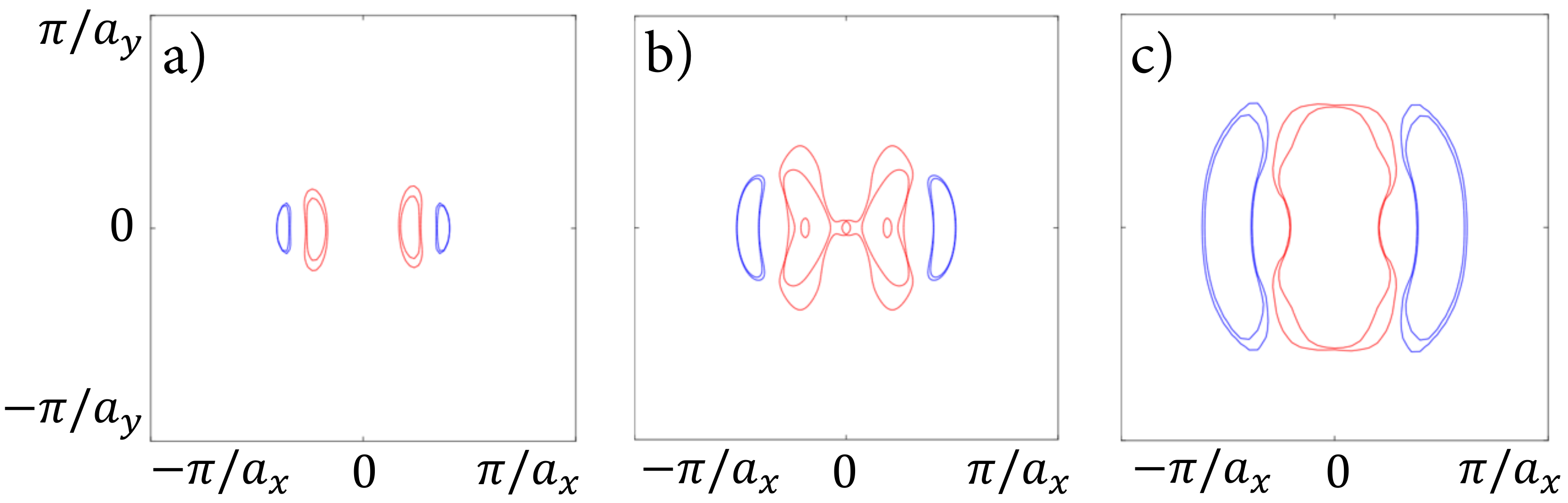}
	\caption{Candidate Fermi surfaces for WTe$_2$, with electron (hole) pockets in blue (red). 
		(a) Highly anisotropic pockets, (b) large pockets and (c) very small pockets, corresponding to a semimetallic state. We find that the effective electron-electron mean free path does not depend sensitively on the choice of Fermi surface.}
	\label{fig:fig2}
\end{figure}

It has been argued~\cite{Vool2021,AharonSteinberg2022} that WTe$_2$ cannot be in the hydrodynamic regime below $20\mathrm{K}$. However, estimates for the effective electron-electron mean free path $\ell_{ee}$ may vary considerably depending on the employed band structure model and other details of the calculational approach. 
Therefore, before invoking a surface mechanism we briefly comment that bulk scattering is unequivocally to weak to matter in these mesoscopic devices.
To this end, consider the electronic self-energy for three qualitatively different, but realistic candidate band structures of the 3-dimensional phase of WTe$_2$. 
The proposed Fermi surfaces describe a (i) a Weyl semimetal phase with very small pockets, (ii) phase with highly anisotropic pockets and (iii) a relaxed phase with large pockets (Fig.~\ref{fig:fig2}).
Employing standard methods to calculate the imaginary part of the self-energy due to the screened Coulomb interaction~\cite{DellAnna2006} and using a fine momentum grid (cf. supplementary information), we obtain the estimates $\ell_{ee}^{(i)}=19 \mathrm{\mu m}$, 
$\ell_{ee}^{(ii)}= 22 \mathrm{\mu m}$ and $\ell_{ee}^{(iii)}= 155 \mathrm{\mu m}$ at $T=4.5\mathrm{K}$. Therefore, we conclude that details of the calculation, and differing starting assumptions about the qualitative shape of the Fermi pockets do not substantially affect the electron-electron scattering rate, soundly excluding bulk mechanisms as the source of hydrodynamic flow. Note that Ref.~\cite{Vool2021} additionally considered phonon-assisted electron-electron interactions, but they are likewise to weak at low T.

\subsection{Boundary scattering model}
Due to the nature of para-hydrodynamic transport regime, the derivation cannot rely on a hydrodynamic treatment, but has to start from a Boltzmann transport approach.
In the following, we consider the nearly ballistic flow in a three-dimensional rectangular geometry where $d\ll w\ll \ell$ (cf. Fig.~\ref{fig:schematics}).
In the kinetic approach, the distribution function is denoted by $f(\mathbf{r},\mathbf{k})$ at real-space position $\mathbf{r}$ and for momentum $\mathbf{k}$ on a spherical Fermi surface. While the material in consideration has a complicated, non-spherical Fermi surface, the relevant Fermi surface quantity in the Boltzmann equation is the Fermi velocity, which is indeed relatively isotropic.
The Boltzmann equation in the steady state is
\begin{align}
\mathbf{v}_F\cdot\nabla_{\mathbf{r}} f
+e\mathbf{E}\cdot\nabla_{\mathbf{k}} f&=\mathcal{I}_{0}(f)+\mathcal{I}_b(f).
\label{eq:boltzmann}
\end{align}
Here, $\mathbf{v}_F$ is the Fermi velocity, $e$ is the electron charge, $\mathbf{E}=(E_x,0,0)$ is the electric field and $\mathcal{I}_0(f)=|\mathbf{v}|(f-f_0)/\ell$ is the bulk collision integral in relaxation time approximation. $\mathcal{I}_b(f)$ is the collision integral due to boundary scattering from the top and bottom surfaces.
Because the thickness $d$ of the flake is much smaller than its width $w$, we can neglect the spatial dependence along the width of the channel. We note that this approximation becomes exact when the in-plane boundaries at $y=\pm w/2$ are completely specular, in which case the distribution function is $y$-independent. We parametrize $f(\mathbf{r},\mathbf{k})-f_0=A h(z,\theta) \cos\theta\cos\phi$, choosing $-d/2\leq z\leq d/2$ along the third dimension, $-\pi/2\leq\phi\leq \pi/2$ as the angle in the plane, and  $-\pi<\theta\leq \pi$ as the out-of-plane angle, with $\theta=\phi=0$ pointing along $+x$ (cf. Fig.~\ref{fig:schematics}). The dimensionless coefficient $A=-(\partial_\epsilon f_0)e E_x\ell$ is chosen such that the solution $h$  becomes normalized ($h(z,\theta)=1$) when boundary scattering is absent, associated with the bulk current density $j_0=\tfrac{e v_F}{32\pi} \int d^3 \mathbf{k} A$. 

We now construct $h(z,\theta)$ which solves Eq.~\eqref{eq:boltzmann} for boundary scattering from a rough surface.
The microscopic process of boundary scattering has been studied since many decades~\cite{Soffer1967,Okulov1975,Falkovsky1983,Moraga1987,Einzel1990,Akhmerov2008,Kiselev2019,Raichev2022}.
It is common to rewrite the boundary collision integral in terms of boundary conditions using a specularity parameter $R_\theta$ which may depend on the angle of incidence $\theta$. Here, $R_\theta=0$ corresponds to completely diffuse scattering, while $R_\theta=1$ is fully specular. Using this parameter, one can express the reflected part of the distribution function $h(d/2,-|\theta|)$ at the top surface in terms of the incident one as $h(d/2,-|\theta|)=R_\theta h(d/2,|\theta|)$, and vice versa at the bottom surface it is $h(-d/2,|\theta|)=R_\theta h(-d/2,-|\theta|)$.
However, using a true 2d scattering cross section to describe a boundary scattering event, one has to go back to the full distribution function, expressing the reflected $f^>$ in terms of the incident $f^<$ by an integral condition~\cite{Falkovsky1983}
\begin{align}
    f^>(\mathbf{k})
    &=f^<(\mathbf{k})+
    k_z\!\!\int_{FS'}\!\!\!\!\!\!\! d^2k' k'_z
    W(\mathbf{k}-\mathbf{k'})[f^<(\mathbf{k'})-f^<(\mathbf{k})]
    \label{eq:generalboundary}
\end{align}
where $W(\mathbf{k})$ is the correlation function of the surface scattering potential, and the integral runs over the half-sphere (FS') of the Fermi surface which corresponds to trajectories incident to the boundary.
We henceforth employ a generic Gaussian-correlated scattering potential~\cite{Falkovsky1983}, defined as $W(\mathbf{k})=\pi a^2b^2e^{-\mathbf{k}^2b^2/4}$, with potential depth $a$ and correlation length $b$.

Historically, Eq.~\eqref{eq:generalboundary} has been solved in two limiting cases, for grazing angles $\theta\approx 0$ and for steep angels of incidence, $\theta\approx \pi/2$~\cite{Falkovsky1983}. 
For grazing angles, the distribution function changes more rapidly than the scattering cross section $W$ and one obtains a standard form of the boundary in terms of the angle-dependent specularity $(1-q|\theta|)$, with parameter $q=4\sqrt{\pi}\Gamma(\tfrac{3}{4})a^2k_F^{3/2}/\sqrt{b}$.
Conversely, for steep angles the scattering potential $W$ changes faster than the distribution function. Using a saddle-point approximation therefore yields a Fokker-Planck equation in the angle of the form $h(d/2,-|\theta|)=\hat O_\theta h(d/2,|\theta|)$ with the operator of angular diffusion on the Fermi surface being $\hat O_\theta=Q\sin^2\theta((2\cot\theta-\tan \theta)\partial_\theta + \partial_\theta^2)$, where  $Q=8a^2/b^2$.
To our knowledge no attempts have been made to treat the scattering for both limits, $\theta\approx 0$ and $\theta\approx \pi/2$ in a unified framework.
However, to capture the para-hydrodynamic behavior we seek a solvable description which holds for all angles of incidence.
Indeed, as we demonstrate next, the description for general angle $\theta$ is absolutely vital to correctly capture the parametric dependencies of the para-hydrodynamic flow. 

In the spirit of Matthiessen's rule, we propose to add both scattering limits as two different types of scattering processes, which yields for the distribution function at the upper boundary of the slab the boundary condition
\begin{align}
    h(d/2,-|\theta|)
    &=(R_\theta+\hat O_\theta)h(d/2,|\theta|).
    \label{eq:newboundary}
\end{align}
Here, we introduced the specularity parameter $R_\theta=(1-q\sin|\theta|)$, which is the periodic extension of the previously mentioned specularity at small angles. 
Note that both scattering types can be combined safely because for $\theta=0,\pi$, the momentum-relaxing scattering vanishes ($R_0=1$), meaning that the distribution function has no discontinuities anywhere.
Equation~\eqref{eq:newboundary} presents the key innovation for a unified treatment of boundary scattering.

The ordinary differential equation~\eqref{eq:newboundary} exhibits several favorable properties.
Firstly, solutions for $q=0$ and for any $Q>0$ are non-dissipative, with distribution function $h(z,\theta)\equiv 1$, corresponding to a bulk current profile without any stresses.
This can be understood as follows. $Q$ parameterizes the relative importance of angular diffusion due to scattering from the boundaries. However, without any momentum loss (i~e. unless $q>0$), the angular diffusion of momentum will redistribute momenta equally into higher and lower momentum states, thereby conserving momentum exactly.
Secondly, the differential equation becomes stiff both at $\theta=0$ and $\theta=\pi/2$, both of which constitute singular points. Therefore, the resulting distribution function at either point singularly depends on the initial conditions at the respective other point. Indeed, we find that for $Q\gg q$ the solution retains a singular dependence on $q$ for all angles, even though $R_\theta$ is dominant over $\hat O_\theta$ only for shallow angles smaller than $d/\ell$.

\subsection{Intermediate regime of para-hydrodynamics}
Using a symmetric parametrization in terms of the characteristics of the motion~\cite{Holder2019}, the solution of the Boltzmann equation can be written as
\begin{align}
    h(z,\theta)&=
    e^{-z\csc \theta/\ell} c(\theta)+(1-e^{-z\csc \theta/\ell}).
    \label{eq:solboltzmann}
\end{align}
$c(\theta)$ is symmetric in $\theta$ and is determined by the modified boundary scattering condition. Inserting the general solution of Eq.~\eqref{eq:solboltzmann} into \eqref{eq:newboundary}, one obtains for $0<\theta<\pi/2$
\begin{align}
    e^{\alpha \csc \theta} c(\theta)+h_{in}^-
    &=(R_\theta+\hat O_\theta)(e^{-\alpha \csc \theta} c(\theta)+h_{in}^+)
    \label{eq:fullbound}
\end{align}
where $\alpha=d/2\ell$ and $h_{in}^\pm= (1-e^{\mp\alpha \csc \theta})$.
Solving Eq.~\eqref{eq:fullbound} is rather involved (cf. supplementary information). 
Typical solutions for $q=1$ and several values of $Q$ are shown in Fig.~\refsub{fig:fig3}{a}. 
For $Q\approx 0$, the distribution function quickly decays with increasing angle, meaning that the current originates almost exclusively from long-lived trajectories in the forward direction ($\theta=0$). This is the expected behavior for ballistic flow. In contrast, for $Q>1$, the profile changes qualitatively, with the trajectories at grazing angles being strongly suppressed, while the distribution function becomes a cosine (corresponding to $c(\theta)=\mathrm{const.}$) for all other angles. This shape of the distribution function resembles a hydrodynamic distribution function. 
The flat part is characterized by the asymptotic value $c_1=c(\pi/2)$, which by Taylor expansion in $\theta\approx\pi/2$ evaluates to
\begin{align}\label{eq:mainsolution}
    c_1
    &=
    1-\frac{qe^\alpha-2Qc''(\tfrac{\pi}{2})}
    {e^{2\alpha}-1+q+2Q\alpha}
    \\
    &\approx 
    \frac{2 Q c''(\tfrac{\pi}{2})+ (2 - q + 2 Q)\alpha }{q + 2 (1 + Q) \alpha }\qquad\text{for $\alpha\to 0$},
\end{align}
where $c''(\pi/2)=\tfrac{1}{2}\partial^2_\theta c(\theta)|_{\theta=\pi/2}$.
In the ballistic regime ($Q=0$), we find numerically that $c''(\pi/2)\propto\alpha$, and thus also $c_1\propto\alpha$, which is subleading compared to the forward trajectories at $\theta\approx0$, which  contribute to the current at order $\mathcal{O}(\sqrt{\alpha})$~\cite{Falkovsky1983}. In contrast, for $Q>0$ we find that $c''(\pi/2)\propto\alpha\log\alpha^{-1}$. 
The weight of trajectories with steep angles is therefore substantially increased compared to the forward trajectories, which are in turn suppressed and contribute to the current only at order $\mathcal{O}(\alpha)$. For small values of $\alpha$, the enhancement of the flat part of the distribution function compared to the forward (ballistic) trajectories is therefore large enough so that a para-hydrodynamic regime emerges (cf.~Fig.~\ref{fig:schematics}).

Since the logarithmic enhancement makes it impossible to construct the size of $c_1$ from local properties around $\theta=\pi/2$, we now focus on the limit where $\alpha\to 0$.
To leading order in $\alpha$, in terms of the variable $s=\sin\theta$, Eq.~\eqref{eq:fullbound} can be expanded as
\begin{align}
&s_\perp^2\left(2+2Q s^2-qs\right)\alpha
=
s_\perp^2\left[(2+2Qs^2)\alpha+qs^2\right]c(s)
\notag\\&
-\frac{Q s^2}{s_\perp^2} \left[s \left(3 s^4-4 s^2+2\right)+2 \alpha  s_\perp^4\right] c'(s)-Q s^3 c''(s)
\end{align}
where $s_\perp=\sqrt{1-s^2}$. This equation does no longer contain any essential singularities for shallow $\theta$ and is much easier to handle numerically. 
Based on the analytical structure of $c''(\pi/2)$ in terms of higher derivatives, and using the numerical solution as a reference, we find by fitting that $c''(\pi/2)\approx \alpha\log(0.06/\alpha)/(0.4 q + 0.5 Q)$.

\subsection{Effective mean free paths}
The crossover from ballistic to bulk hydrodynamic flow is well studied in two-dimensional electron fluids~\cite{Alekseev2016,Shytov2018,Holder2019,
deJong1995,Scaffidi2017,Kiselev2019,Ledwith2019,
Afanasiev2021,Li2021}, 
and is typically described using a 
dual relaxation time approximation for the scattering integral in terms of the two length scales $\ell_{mr}$ and $\ell_{mc}$.
However, it has been shown~\cite{Holder2019,Holder2019a} that the nature of the transport regime can also be reconstructed from an inspection of the distribution function.
Namely, the distribution is smooth and relatively flat in the hydrodynamic regime ($\ell_{mc}\ll w\ll \ell_{mr}$), while in the ballistic regime ($w\ll \ell_{mr},\ell_{mc}$), scattering from the boundaries makes the distribution function strongly angle dependent. 
This can be more formally restated by considering the angular harmonics of $h(\theta)$. 
Keeping only the first and second term in such an expansion, the Boltzmann equation simplifies to the Stokes-Ohm equation~\cite{Holder2019}, i.e. the distribution function can be obtained in a hydrodynamic description.
On the other hand, if higher angular harmonics are present in the distribution function, this indicates the presence of additional long-lived modes in the flow, as it would be expected for a ballistic distribution function.

\begin{figure}
	\centering
	\includegraphics[width=\columnwidth]{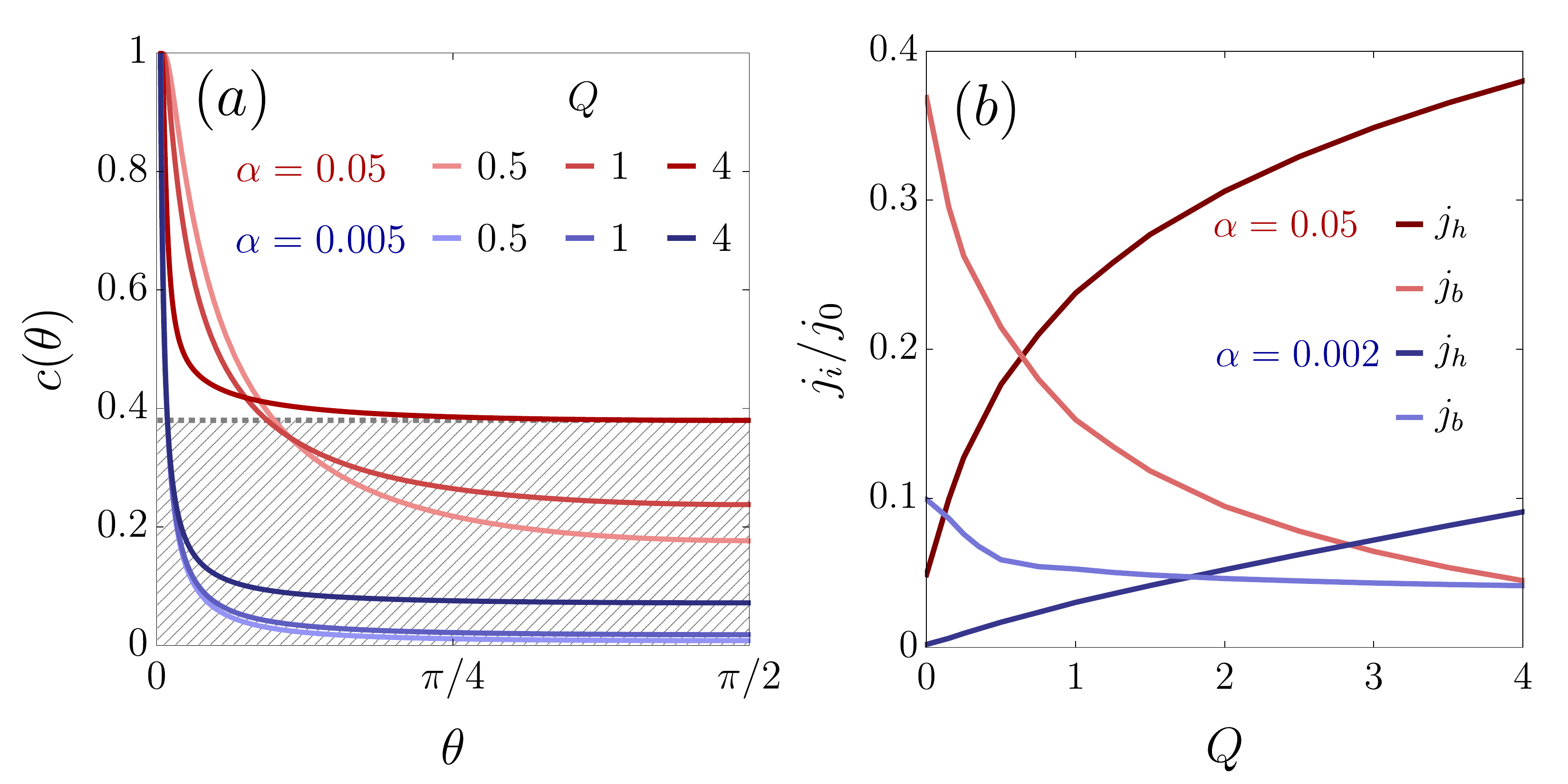}
	\caption{Normalized distribution function and currents in the two-fluid picture. (a) The distribution function $c(\theta)$ in the para-hydrodynamic regime, for $q=1$, three different values of $Q$ and two $\alpha$.
		With increasing $Q$, more weight is accumulated at the steep trajectories away from $\theta=0$. 
		For $\alpha=0.05$, $Q=4$, the asymptotic value $c(\pi/2)=c_1$ is indicated by a gray dashed line. In the two-fluid approximation the current $j_h$ stemming from the area below $c_1$ (gray, hatched) is compared against the current $j_b$ induced by the rest of the non-equilibrium distribution.
		(b) Comparison of the para hydrodynamic ($j_h/j_0$) and ballistic ($j_b/j_0$) current density contributions in the two-fluid approximation for two values of $\alpha$, where $j_0$ denotes the bulk current density.
		For all $\alpha\ll 1$, the contribution of the para-hydrodynamic current increases with $Q$, quickly overtaking the ballistic contribution. 
	}
	\label{fig:fig3}
\end{figure}

In the present case, different parts of the distribution function at respectively shallow or steep angles resemble either the ballistic or the hydrodynamic situation. 
We therefore propose a two-fluid approximation (cf. Figs.~\ref{fig:schematics},\ref{fig:fig3}), whereby we decompose the distribution function $h(\theta)$ into a constant part which constitutes a hydrodynamic current density $j_h= c_1 j_0$, while the remaining strongly angle dependent parts constitute a ballistic current density $j_b=j-j_h$.

As shown in Fig.~\refsub{fig:fig3}{b}, at $Q=0$ it is $j_h<j_b$, but upon increasing $Q$ there is a crossover into a regime with $j_h>j_b$.
In the latter regime, one can immediately infer $\ell_{mr}$ and $\ell_{mc}$ from the flat part of the distribution function which creates the dominant contribution $j_h$ to the current density. To this end, using the reduced current density we write for the effective momentum relaxing mean free path,
\begin{align}
    \frac{\ell\ell_{mr}}{(\ell+\ell_{mr})}
    &=
    \frac{\ell}{\pi}\int_{-\pi}^\pi
    \cos^2\theta
    c(\theta)
    \approx \ell c_1.
\end{align}
The flatness of the distribution function furthermore suggests that the total scattering rate is large and completely dominated by momentum conserving processes.
$\ell_{mc}$ is therefore expected to approach its maximally possible value, which for a flat distribution is entirely determined geometrically by the normalized travel distance between two successive scatterings from the top and bottom surfaces. In other words, we can estimate that $\tfrac{1}{\ell_{mc}}\approx \tfrac{1}{d}\tfrac{1}{2\pi}d\theta \int_{0}^{\pi}\sin|\theta|$, which yields $\ell_{mc}=\pi d$.

Using these values for $\ell_{mr}\ll\ell$ and $\ell_{mc}\ll\ell$, the effective Gurzhi parameter for the para-hydrodynamic flow in thin sheets becomes \begin{align}\label{eq:Gurzhi}
    D'&=\tfrac{1}{2}\sqrt{\tfrac{\ell\ell_{mr}}{\ell+\ell_{mr}}
\tfrac{\ell\ell_{mc}}{\ell+\ell_{mc}}}
\approx\tfrac{\sqrt{\pi}}{2}\sqrt{c_1 d \ell }.
\end{align}
In the experiment, the WTe$_2$ samples had the properties $d=48\mathrm{nm}$, $\ell=20\mathrm{\mu m}$, 
$\ell_{mr}=530\mathrm{nm}$ and $D=155\mathrm{nm}$. 
Using as input the experimental values for $d$ and $\ell$, Eq.~\eqref{eq:Gurzhi} yields $D'=145\mathrm{nm}$ and $\ell_{mr}=560\mathrm{nm}$ for the choice $(q,Q)=(1,4)$, while it is $D'=143\mathrm{nm}$ and $\ell_{mr}=545\mathrm{nm}$ upon choosing $(q,Q)=(0.8,2)$. 
This indicates that the kinetic theory is not strongly sensitive to the precise value of $q$ and $Q$, and it can explain the experimental findings well for a reasonable range of boundary scattering parameters. 
Importantly, this versatility implies that para-hydrodynamic flow emerges generically as long as there is a large scale separation between sheet thickness $d$ and bulk mean free path $\ell$, and  does not require extensive fine-tuning.
We remark that $q$ and $Q$ are in turn only weakly dependent on the microscopic scattering parameters $a$ and $b$. For example, using $k_F=1\mathrm{nm^{-1}}$, we obtain $a=0.27\mathrm{nm}$, $b=0.37\mathrm{nm}$ for $(q,Q)=(1,4)$, and $a=0.26\mathrm{nm}$, $b=0.51\mathrm{nm}$ for $(q,Q)=(0.8,2)$. These estimates correspond to boundary roughness at the lattice scale and are indeed reasonable for cleaved samples~\cite{Woods2016}.

\section{Discussion}
We derived a microscopic boundary scattering model which can explain para-hydrodynamic flow in the absence of strong electron-electron scattering. 
As the main characteristics of the new regime, we identified singular points in the resulting boundary condition, which lead to a different thickness dependence of the current density that onsets in the presence of small angle scattering from the boundary.
Our findings constitute a new type of ballistic-to-hydrodynamic crossover, 
where the three-dimensional problem is microscopically ballistic, but the in-plane components of the flow velocity exhibit relaxation properties which are indistinguishable from viscous flows.
Our results indicate that the phenomenology which was previously suggested to govern the hydrodynamic-to-ballistic crossover is not universal. 
Since para-hydrodynamic flow exclusively emerges in the presence of angular diffusion from short range correlated disorder, conventional approaches which rely solely on a reflectivity coefficient have not been able to capture this mechanism~\cite{deJong1995,Alekseev2018,Svintsov2018}.
We note that small angle scattering can also appear from bulk scattering, in which case it typically leads to a ratio $\ell_{mr}/\ell_{mc}=4$~\cite{Ledwith2019a,Nazaryan2021}. For small angle boundary scattering we instead found $\ell_{mr}=c_1\ell$ and $\ell_{mc}=\pi d$, which yields a scale-dependent ratio $\ell_{mr}/\ell_{mc}\propto\log(\ell/d)$ that becomes large enough to support hydrodynamic phenomena for very large ratios $\ell/d$.

It would be interesting to find additional signatures of the para-hydrodynamic regime for the channel flow. Since the in-plane current density in narrow channels is not suitable for distinguishing between the ballistic and hydrodynamic transport regimes~\cite{Sulpizio2019}, this would probably involve the investigation of the Hall viscosity at finite magnetic fields~\cite{Delacretaz2017,Scaffidi2017} or optical probes.

Since only very few microscopic parameters other than the bulk mean free path enter into our results, we expect that the para-hydrodynamic flow observed in WTe$_2$ is not unique, and that a number of clean three-dimensional compounds, for example Weyl and Dirac semimetals~\cite{Yan2017,Armitage2018} and ultrapure delafossites~\cite{Moll2016,Mackenzie2017}, can exhibit this new transport regime.
The proposed boundary scattering model and the two-fluid approximation used here for extracting the effective mean free paths are generic. Thus, the same methodology can very likely be applied directly to many other three-dimensional materials and also integrated into numerical schemes~\cite{Varnavides2022}.

\section{Methods}
The calculation of the electron-electron scattering rate was done in four steps.
The first step in the process is finding the energy bands and wave functions of the Weyl Semimetal phase of WTe$_2$ within DFT using VASP and Wannier90 \cite{DFT,gradients}. These were evaluated inside the Brillouin zone on a k-mesh of $N_x\times N_y\times N_z=100\times 50\times 7$.
The particle-hole bubble $\Pi$ was calculated with an IR cutoff of $5meV$, which is approximately a tenth of the Fermi energy. The real and imaginary part were calculated separately to avoid numerical errors.
$\Re\Pi(q,\omega)$ and $\Im\Pi(q,\omega)$ were evaluated on the same k-mesh as the bands and wave functions, and on energies between $-11eV$ to $11eV$ with a resolution of $0.055eV$. On energies in between the grid points a linear interpolation was used.
The UV energy cutoff is chosen such that the low temperature self energy corrections are converged, where the main contribution to the relaxation time is by electrons with energy $\epsilon_{nq}$ in band $n$ residing in an energy window of $\epsilon_F-T<\epsilon_{nq}<\epsilon_F+T$ that are scattered by electrons in band $m$ that likewise satisfy $\epsilon_F-T<\epsilon_{mk}<\epsilon_F+T$. Thus, the main contribution to the self energy is generated by electrons whose differences in energies are $\omega=\epsilon_{nq}-\epsilon_{mk}<2T$, while the target temperatures are below $T<0.025eV \ll 11eV$. It is then sufficient to let the band indices run through the first 10 bands above and below the Fermi surface, which corresponds to a maximal energy difference of $4eV$. 

The self energy was calculated with a regularized Bose distribution function, $b(\epsilon)\rightarrow (e^{\beta \epsilon}-1)/((e^{\beta \epsilon}-1)^2+\lambda)$ (cf. supplementary information).
A good choice for the regulator is $\lambda=10^{-4}$, such that the pole at zero frequency is cut off at $b(\epsilon)<10^4$.

We confirmed the convergence of our results by comparing various grid resolutions. Firstly, two k-meshes were chosen at sizes $200\times 50\times 3$ and $100\times 100\times 3$. The difference in the self-energy between both grid resolutions is less than 2\%. Additionally, the frequency resolution was also tested by increasing the resolution to $0.02eV$, which yielded a change of less than 1\%. Finally, the bosonic occupation number cutoff was increased to $\lambda=10^-6$, which yielded a change of less than 1 percent. The sensitivity of our results against small changes of the Fermi level was also tested. On two points separated by $0.02eV$ the results changed by about 36\% which is significant but shows that the order of magnitude is robust to small changes in the chemical potential. 
A calculation with a different number of bands included into the correction of the photon propagator showed no relevant changes.

\section{Data Availability}
The data that support the findings of this study are available from the corresponding author upon request.

\section{Acknowledgments}
We thank 
J.~S.~Hofmann
and
E. Zeldov
for fruitful discussions.
B.Y.\ acknowledges the financial support by the European Research Council (ERC Consolidator Grant No. 815869, ``NonlinearTopo'') and Israel Science Foundation (ISF No. 2932/21). 

\section{Author contributions}
T.H. conceived the project. T.H. and Y.W. did the analytical and numerical calculations for the boundary model. Y.W. did the DFT and
e-e scattering rate calculations. Y.W., A.A.-S., B.Y. and T.H. analyzed the results. 
T.H. wrote the manuscript with the input from all other authors.
B.Y. supervised the project.

\section{Competing interests}
The Authors declare no competing interests.

\renewcommand{\thefigure}{\arabic{figure}}
\renewcommand{\figurename}{Supplementary Figure}
\renewcommand{\tablename}{Supplementary Table}
\setcounter{figure}{0}
\setcounter{equation}{0}
\renewcommand{\theequation}{S\arabic{equation}}

\clearpage

\onecolumngrid

\begin{center}
	\textbf{\large Supplementary material for: \strut\\ ``Para-Hydrodynamics from weak surface scattering in thin WTe$_2$ flakes'' }\\[5pt]
	\begin{quote}
		{\small 
    		In this supplementary information we present the calculation of the electron-electron mean free path, derive the operator of angular diffusion, and detail the solution of the differential equation for boundary scattering.}\\[20pt]
	\end{quote}
\end{center}

\section{Electron-electron collision rate from the screened Coulomb interaction}

To quantify the strength of the interaction we focus the study on the relaxation length, $l_{ee}$, the mean free path for electron-electron collisions. In Fermi liquid theory normal metals have a relaxation length with a $T^{-2}$ temperature dependence. 
We emphasize that in WTe$_2$, the assumption of a $T^{-2}$-dependence is not justified \emph{a priori}. Instead, it is necessary to calculate $l_{ee}$ down to low temperatures in order to extract a reliable temperature dependence. 
In our previous calculation~\cite{AharonSteinberg2022}, we indeed found that the low temperature limit approaches a $T^{-2}$ behavior, but such happens only at energies much smaller than the Fermi level, which in WTe$_2$ is unusually low at $E_F\approx 150\mathrm{K}$.
Since the Fermi energy, and the Fermiology of the Fermi surface play a prominent role in WTe$_2$~\cite{Vool2021}, and since our previous estimate for $\ell_{ee}\approx 10\mathrm{\mu m}$ was only 1.5 orders of magnitude to large to explain the hydrodynamic electron flow, here we repeat our calculation for two other band structure candidates for WTe$_2$. This way, we can make sure that the reported absence of a sufficiently small $\ell_{ee}$ is not an artifact of the specific choice of band structure, but holds invariably for this material.

\subsection{Formalism}
Electron-electron interactions arise from the coupling of the electrons to the electromagnetic field. In 3D, the bare coulomb interaction is given by $U_{bare}(q)=\frac{4\pi e^2}{|q|^2}$. The presence of electrons modifies the bare interaction, such that the renormalized interaction $W$ becomes
\begin{equation}
    U(q,\omega)=\frac{4\pi e^2}{|q|^2-4\pi e^2 \Pi(q,\omega)}.
\end{equation}
Here, $\Pi(q,\omega)$ is the particle-hole susceptibility of the electromagnetic field, given by the one-loop integral
\begin{equation}
    \Pi(q,\omega)=\sum_{i,j}\int \frac{d^3k}{(2\pi)^3}
    |\rho_{i,k,j,k+q}|^2\frac{n(\xi_{i,k})-n(\xi_{j,k+q})}{\xi_{i,k}-\xi_{j,k+q}+\omega-i\eta}.
\end{equation}
$i,j$ are band indices, $n(\epsilon)$ is the Fermionic occupation number, $\xi=\epsilon-\epsilon_F$ and $|\rho_{i,k,j,k+q}|^2=|\langle{u_{i,k}}e^{iqr}|u_{j,k+q}\rangle|^2$ is the exchange density.
At finite temperature the electronic self-energy due to the Fock diagram is
\begin{equation}\label{eq:selfe}
    \Im\Sigma(q,\epsilon_{nq})=-\pi\sum_m\int \frac{d^3k}{(2\pi)^3}|\rho_{n,q,m,k}|^2
    (b(\epsilon_{mk}-\epsilon_{nq})+n(\epsilon_{mk}))\Im U(k-q,\epsilon_{mk}-\epsilon_{nq}),
\end{equation}
where $b(\epsilon)$ is the bosonic occupation number.
The electron-electron relaxation rate of a state is defined in terms of the self energy as
\begin{equation}
    [\tau_{ee}(q,\epsilon_{nq})]^{-1}=\frac{2}{\hbar}\Im\Sigma(q,\epsilon_{nq}).
\end{equation}
The average relaxation rate is then the average of the relaxation rate on the Fermi surface, normalized by the total density of states:
\begin{equation}
    \langle\tau_{ee}^{-1}\rangle_{FS}=\frac{\sum_n\int \frac{d^3k}{(2\pi)^3}[\tau_{ee}(q,\epsilon_{nq})]^{-1}\frac{\partial f_{nq}}{\partial \epsilon_{nq}}}{\sum_n\int \frac{d^3k}{(2\pi)^3}\frac{\partial f_{nq}}{\partial \epsilon_{nq}}}
\end{equation}
Finally, the relaxation length is given as the inverse of the relaxation rate,
\begin{equation}\label{eq:lee}
    l_{ee}=v_F\langle\tau_{ee}^{-1}\rangle_{FS}^{-1}.
\end{equation}

\subsection{Candidate band structures}

\textbf{Weyl semimetal candidate band structure} The chemical potential resides at $\mu=7.22eV$, the lattice parameters are $a_x = 0.3483nm, a_y = 0.6265nm, a_z = 1.4043nm$.
The resulting parameters are listed in Supplementary Table~\ref{tab:weyl}.

\begin{table}[h]
    \centering
    \begin{tabular}{ |c|c|c|c|c|c|c|c| }
    \hline
         $n$ & $\epsilon_f[eV]$ & $v_F[10^5m/s]$ & $(k_F)_x[nm^{-1}]$ & $(k_F)_y[nm^{-1}]$ & $(k_F)_z[nm^{-1}]$ & $g(\epsilon_f)[eV^{-1}nm^{-3}]$ & $n(\epsilon_f)[nm^{-3}]$ \\
         \hline
         55 & 0.0496 & 1.09 & 0.60 & 0.63 & 0.54 & 2.1655  & 0.0631\\
         56 & 0.0834 & 1.327 & 0.61 & 0.92 & 0.54 & 2.9942 & 0.1056\\
         57 & 0.0691 & 2.35 & 0.34 & 0.60 & 0.93 & 1.0546  & 0.0498\\
         58 & 0.0679 & 2.32 & 0.30 & 0.55 & 0.89 & 0.9026  & 0.0393\\
    \hline
    \end{tabular}
    \caption{Details of the Weyl semimetal phase.}
    \label{tab:weyl}
\end{table}

\textbf{Relaxed candidate band structure}
The chemical potential sits at $\mu=8.5083$. This band structure is motivated by Ref.~\cite{Vool2021}, 
with lattice parameters $a_x = 0.3460nm, a_y = 0.6200nm, a_z = 1.3090nm$. Compared to the other band structures, this corresponds to a compression along the $k_z$-direction. The position of the atoms inside the lattice was subsequently optimized to minimize the total energy. The result is shown in Supplementary Table~\ref{tab:relaxed}.

\begin{table}[h]
    \centering
    \begin{tabular}{|c|c|c|c|c|c|c|c|c|}
    \hline
         $n$ & $\epsilon_f[eV]$ & $v_F[10^5m/s]$ & $(k_F)_x[nm^{-1}]$ & $(k_F)_y[nm^{-1}]$ & $(k_F)_z[nm^{-1}]$ & $g(\epsilon_f)[eV^{-1}nm^{-3}]$ & $n(\epsilon_f)[nm^{-3}]$ \\
         \hline
         55 & 0.1944 & 2.91 & 1.23 & 1.48 & 1.27 & 1.4147  & 0.2752\\
         56 & 0.2074 & 2.95 & 1.35 & 1.74 & 1.28 & 1.7539 & 0.3501\\
         57 & 0.2565 & 2.80 & 0.73 & 1.48 & 1.11 & 2.6862  & 0.3231\\
         58 & 0.2307 & 3.03 & 0.64 & 1.30 & 1.15 & 1.9589  & 0.2338\\
         \hline
    \end{tabular}
    \caption{Details of the relaxed band structure.}
    \label{tab:relaxed}
\end{table}

\textbf{Candidate band structure based on hybrid functionals}
Here, the chemical potential is $\mu=6.965eV$. The chosen lattice parameters are $a_x = 0.3483nm, a_y = 0.6265nm, a_z = 1.4043nm$. 
Supplementary Table~\ref{tab:third} lists the band structure parameters obtained when using hybrid functionals in the self consistent stage of the calculation in order to better account for exchange.

\begin{table}[h]
    \centering
    \begin{tabular}{|c|c|c|c|c|c|c|c|c|}
    \hline
         $n$ & $\epsilon_f[eV]$ & $v_F[10^5m/s]$ & $(k_F)_x[nm^{-1}]$ & $(k_F)_y[nm^{-1}]$ & $(k_F)_z[nm^{-1}]$ & $g(\epsilon_f)[eV^{-1}nm^{-3}]$ & $n(\epsilon_f)[nm^{-3}]$ \\
         \hline
         55 & 0.0669 & 2.57 & - & - & - & 0.2158  & 0.0065\\
         56 & 0.0801 & 2.49 & - & - & - & 0.2923 & 0.0101\\
         57 & 0.0430 & 2.95 & - & - & - & 0.3450  & 0.0120\\
         58 & 0.0419 & 2.68 & - & - & - & 0.1587  & 0.0040\\
         \hline
    \end{tabular}
    \caption{Details of the band structure obtained from hybrid functionals. 
    }
    \label{tab:third}
\end{table}

We briefly comment on consistency checks using the density of states, and why they are less relevant in the present calculation.
For example, the total density of states of the relaxed phase is $7.81 eV^{-1}nm^{-3}$ and according to the calculation of the real part of the susceptibility at $\omega=0.01eV$ it is $6.47 eV^{-1}nm^{-3}$, which agrees within 20\% with the total density of states.
Likewise, the total density of states of the Weyl semimetal phase is $7.12 eV^{-1}nm^{-3}$ and according to the calculation of the real part of the susceptibility at $\omega=0.01eV$ it is $4.27 eV^{-1}nm^{-3}$, which is further apart, with a 67\% difference.
As the main reason we suspect the more sensitive dependence on $\omega$ in the Weyl semimetal band structure. However, for a faithful estimate of the self-energy this effect is not relevant because we are predominantly interested in the imaginary part of $U$, which becomes zero exactly at $\omega=0$.

\subsection{Results for the electron-electron mean free path}

\begin{figure}
    \centering
    \includegraphics[width=.45\columnwidth]{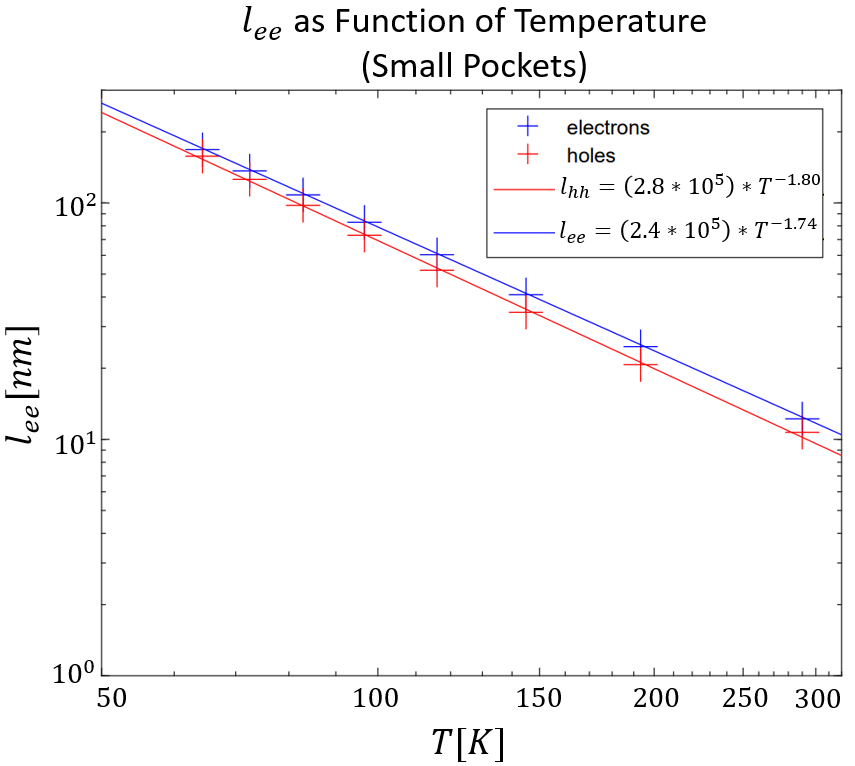}
    \includegraphics[width=.45\columnwidth]{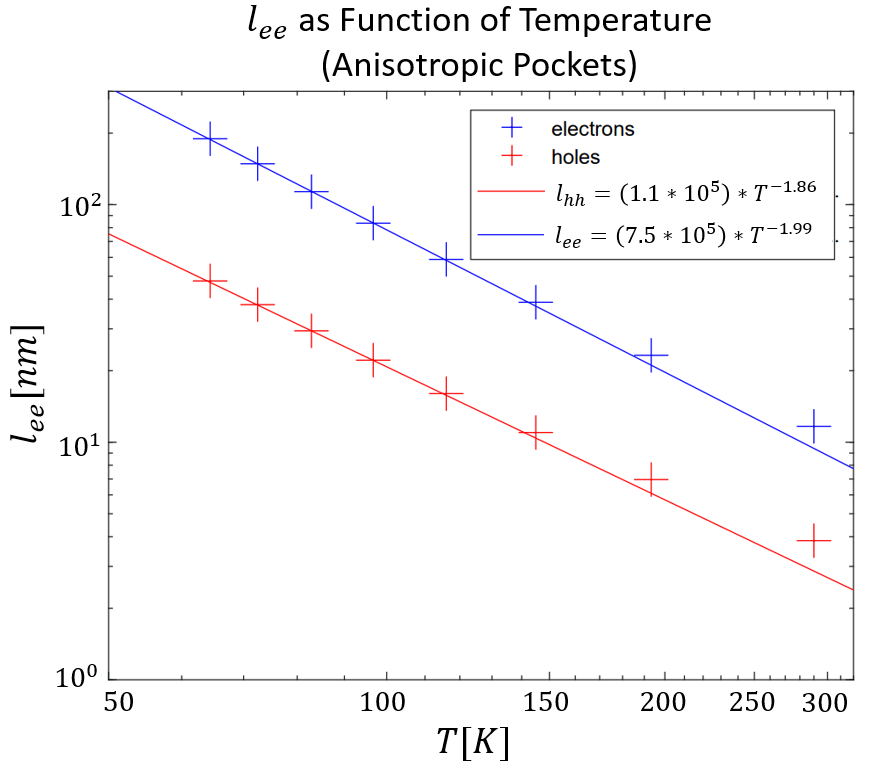}
    \includegraphics[width=.45\columnwidth]{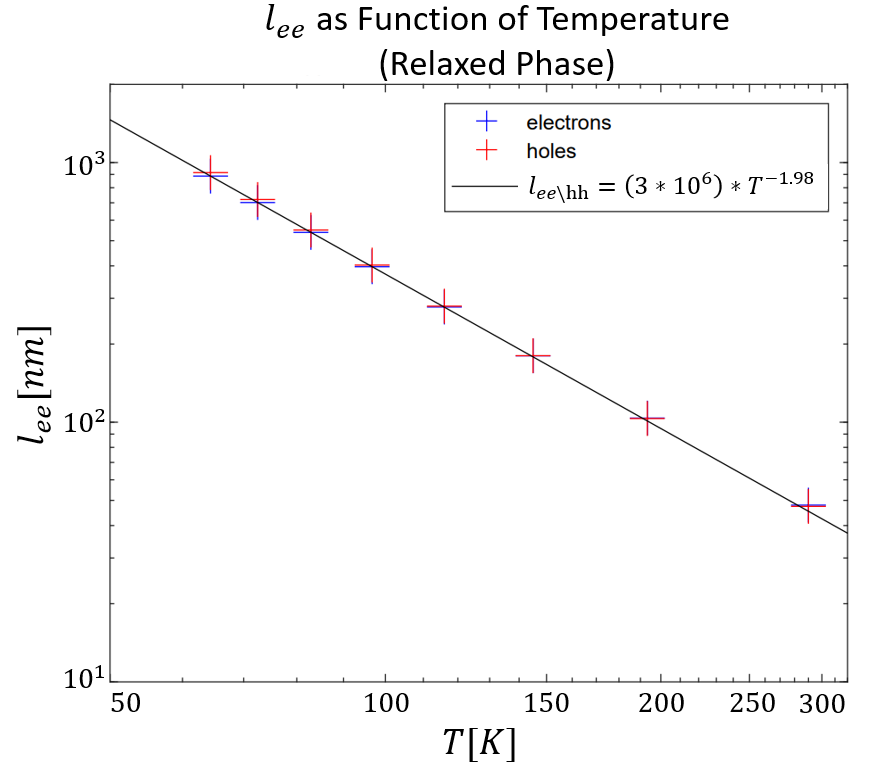}
    \caption{$l_{ee}$ as function of temperature for the three band structures. (a) Small pockets: due to the compensation of electrons and holes and the similar velocities of the bands, $l_{ee}$ and $l_{hh}$ are similar, we can see a good fit to a power law, but with a weaker dependence on temperature than inverse square. (b) Anisotropic pockets: the Fermi level was chosen in a way that the compensation of holes and electrons is not perfect, more electrons than holes. This is reflected both in an effective lower Fermi level and lower velocity for the holes, which manifest in the form of $l_{hh}<l_{ee}$. Still, we can see a good fit to a power law, specifically for electrons, a $T^{-2}$. (b) Relaxed phase: resembels a metal and as such fits well to a $T^{-2}$ power law. The relatively large $l_{ee}$ compared to the previous two band structures is mainly due to the large Fermi level, which leads to high screening, reducing the effective interaction.}
    \label{fig:calculatedee}
\end{figure}

The electron-electron mean free path according to Supplementary Eq.~\eqref{eq:lee} and its temperature scaling are shown in Supplementary Fig.~\ref{fig:calculatedee} for all three candidate band structures. Clearly, the low-energy asymptotic scaling onsets at or even slightly above $150\mathrm{K}$ in all three cases. Extrapolating the obtained temperature scaling, we arrive at the low-temperature quoted in the main text.

\section{Differential Operator}
The derivation of the differential operator was first done in Ref.~\cite{Falkovsky1983}, but uses a highly customized notation. For this reason, we reiterate here the steps needed to obtain $\hat O_\theta$.
We start from the general boundary scattering condition in terms of the scattering potential (Eq.~(2) in the main text),
$f^{>}(\bm{k})=f^{<}(\bm{k})+I
$
where the integral is defined as
\begin{align}
I=k_z\int k'_z(f^{<}(\bm{k}')-f^{<}(\bm{k}))W(\bm{k}',\bm{k})    
\end{align}
Recalling that the scattering potential has the explicit form $W(\bm{k}',\bm{k})=W_0 e^{-\frac{|\bm{k}-\bm{k}'|^2}{2\sigma^2}}$, we proceed with a saddle point approximation.
Denoting $g(\bm{k}')=k'_z(f^{<}(\bm{k}')-f^{<}(\bm{k}))$, it is clear that $g(\bm{k})=0$. Expanding up to second order in $\bm{k}'-\bm{k}$, in spherical coordinates on the Fermi surface, it is therefore
\begin{equation}
    g(\bm{k}')=\sum_i [\partial_i g(\bm{k})](k'-k)_i+\frac{1}{2}\sum_{ij}[\partial_i \partial_j g(\bm{k})](k'-k)_i(k'-k)_j
\end{equation}
Inserting the expansion into the integral $I=I_1+I_2$, we obtain
\begin{align}
    I_1 &= W_0 k_z \sum_i [\partial_i g(\bm{k})]\int (k'-k)_i e^{-\frac{|\bm{k}-\bm{k}'|^2}{2\sigma^2}}
\\
    I_2&= \frac{1}{2}k_z W_0\sum_{i,j}[\partial_i^2 g(\bm{k})]\bigg[\int (k'-k)_i (k'-k)_j e^{-\frac{|\bm{k}'-\bm{k}|^2}{2\sigma^2}}\bigg]
\end{align}
In polar coordinates relative to $\bm{k}$, we notice that  $(k'-k)_i$ is anti-symmetric with respect to the transformation $\bm{k}'\rightarrow 2\bm{k}-\bm{k}'$, while $e^{-\frac{|\bm{k}-\bm{k}'|^2}{2\sigma^2}}$ is symmetric. Therefore the integral on the  components that are perpendicular to $\bm{k}$, meaning $i=\theta,\varphi$ vanish. On the other hand, the integral on the radial  direction, does not vanish from this argument, but instead vanishes identically, as $|\bm{k}'|=|\bm{k}|$ is enforced identically on the entire integration range. Thus it is $I_1=0$.

For $I_2$, obviously the same argument holds for the radial direction. It thus suffices to look at directions $i,j$ that are perpendicular to $\bm{k}$. Considering $i\neq j$ next, the integrand is odd in either direction and vanishes due to antisymmetric cancellation. So the only terms that contribute are $i=j$ for the two directions perpendicular to $\bm{k}$,
\[
I_2 = \frac{1}{2}p_z W_0\sum_{i=1,2}[\partial_i^2 g(\bm{k})]\bigg[\int (k'-k)_i^2 e^{-\frac{|\bm{k}'-\bm{k}|^2}{2\sigma^2}}\bigg].
\]
The integral inside the brackets can be solved by noticing that for the two directions perpendicular to $\bm{k}$, the result is the same from rotational symmetry around the $\bm{k}/|\bm{k}|$ axis:
\begin{align}
\int (k'-k)_i^2 e^{-\frac{|\bm{k}'-\bm{k}|^2}{2\sigma^2}}
&=\frac{1}{2}\int (\bm{k}'-\bm{k})^2 e^{-\frac{|\bm{k}'-\bm{k}|^2}{2\sigma^2}}
= -\sigma^2 k_F^2 \frac{\partial}{\partial \alpha}\bigg[e^{-\frac{\alpha k_F^2}{\sigma^2}}\int   e^{\frac{\alpha k_F^2 \cos\theta}{\sigma^2}}
\sin\theta d\theta d\varphi\bigg]|_{\alpha=1}\\
&= 2\pi \sigma^4 \bigg[\frac{2 k_F^2}{\sigma^2}e^{-\frac{2 k_F^2 }{\sigma^2}}-(1-e^{-\frac{2k_F^2 }{\sigma^2}})\bigg]\equiv\frac{2Q}{W_0}
\end{align}
In summary, the integral $I$ evaluates to
\begin{align}
I = I_2 = k_z\sum_{i=1,2}[\partial_i^2 g(\bm{k})]Q 
=
Q k_z \Bigl[\sum_{i=1,2} \partial_i^2\Bigr] (k_z f^<(\bm{k})).
\end{align}
We note that $\partial_i^2$ where $i=\theta,\phi$ is exactly the Laplace-Beltrami operator on the sphere. We can therefore write more formally
\begin{align}
    f^>(\bm{k})=f^<(\bm{k})
    +Q k_z\nabla^2(k_zf^<(\bm{k})).
\end{align}
Using the identity $\nabla^2 (k_z f^<(\bm{k}))=f^<(\bm{k})\nabla^2 k_z+k_z\nabla^2 f^<(\bm{k})+2\nabla k_z\cdot\nabla f^<(\bm{k})$ and noting that $\nabla^2 \bm{k}_z=0$, we recover the form of Ref.~\cite{Falkovsky1983},
\begin{align}
    f^>(\bm{k})=\bigg(1+Q k_z(\nabla k_z\cdot \nabla+2k_z\nabla^2)\bigg)f^<(\bm{k}).
\end{align}
In spherical coordinates, this is
\begin{align}
\nabla k_z\cdot \nabla 
&= (-\sin\theta \hat{\theta})\cdot (\hat{\theta}\frac{1}{k_F}\partial_\theta+\hat{\varphi}\frac{1}{k_F \sin\theta}\partial_\varphi)
=-\frac{\sin\theta}{k_F}\partial_\theta
\\
\nabla^2 &= \frac{1}{k_F^2 \sin\theta}\partial_\theta (\sin\theta \partial_\theta)+\frac{1}{k_F^2 \sin^2\theta}\partial_\varphi^2
\end{align}
Since we consider a thin slab where the thickness $d$ is  much smaller than the width $w$, we can neglect derivatives with respect to the in-plane angle $\phi$, which are expected to only contribute weakly due to the smooth dependence of $f$ on $\phi$.
In term of the reduced distribution function $h(z,\theta)$ the boundary condition therefore simplifies to
\begin{align}
h(d/2,-|\theta|)
&=\Bigl[1+Q\cos\theta\bigl( \cot\theta\partial_\theta (\sin\theta \partial_\theta)-2\sin\theta\partial_\theta \bigr)\Bigr]h(d/2,|\theta|)
\\&=
\Bigl[1+Q\cos^2\theta\bigl((\cot\theta-2\tan \theta)\partial_\theta + \partial_\theta^2\bigr)\Bigr]h(d/2,|\theta|),
\label{eq:suppoperator}
\end{align}
which, after shifting $\theta\to\theta-\pi/2$ and adding the specularity coefficient $R_\theta$ is the form quoted in the main text in Eq.~(3).
While it may look as though Supplementary Eq.~\eqref{eq:suppoperator} contains a divergence at at $\theta=0$, 
we point out that this divergence is of the form $1/\sin\theta$, which is rendered finite upon integration with the Jacobian $k_F^2 \sin\theta$ on the sphere.

\section{Solution of the differential equation for boundary scattering}

The differential boundary condition, Eq.~(5) in the main text, can be rewritten in terms of the variable $s=\sin\theta$, and reads explicitly
\begin{align}
&\left(1-s^2\right) \left[\alpha ^2 Q-s^3 \left(q+2 \alpha  Q-q e^{\alpha /s}\right)+s^2 \left(1-\alpha ^2 Q-e^{\frac{2 \alpha }{s}}\right)\right]
\notag\\
&=
\left(1-s^2\right)\left[\alpha ^2 Q-s^3 (q+2 \alpha  Q)+s^2 \left(1-\alpha ^2 Q-e^{\frac{2 \alpha }{s}}+1\right)\right]c(s)
\notag\\&
+\frac{Q s^2}{1-s^2} \left[2 \alpha  \left(1-s^2\right)^2+s \left(3 s^4-4 s^2+2\right)\right] c'(s)+Q s^4 c''(s).
\label{eq:difffull}
\end{align}
A unique solution for $c(\theta)$ is obtained by imposing smooth continuity for this periodic and even function, i.~e. we demand that $c'(0)=c'(\pi/2)=0$. 
However, since $\theta=0,\pi/2$ are singular points, this way of solving Supplementary Eq.~\eqref{eq:difffull} is impractical, and it is preferable to instead use initial conditions for $c(\theta_0)$ and $c'(\theta_0)$, where $\theta_0\approx 0$ is chosen small enough for the effect of the derivatives in Supplementary Eq.~\eqref{eq:difffull} to become vanishingly small at $\theta_0$. This makes sense because for $\theta\to 0$ Eq.~(5) of the main text can be solved algebraically, yielding
\begin{align}
    c_0(\theta)
    &=
    1-\frac{q e^{\alpha  \csc\theta}}
    {q+\csc\theta \left(e^{2 \alpha  \csc\theta}-\alpha^2Q \cot ^2\theta-1\right)+2 \alpha  Q}.
\end{align}
Then, using initial values $c(\theta_0)=1-(1+\lambda) (1-c_0(\theta))$ and $c'(\theta_0)=(1+\lambda) c'_0(\theta)$, and slowly adjusting the tuning parameter $\lambda\approx 0$ such that the solution does not diverge at $\pi/2$ presents a reliable way to construct physically relevant solutions for $c(\theta)$. A typical iterative solution is presented in Supplementary Fig.~\ref{fig:fitting1}

\begin{figure}
    \centering
    \includegraphics[width=.33\columnwidth]{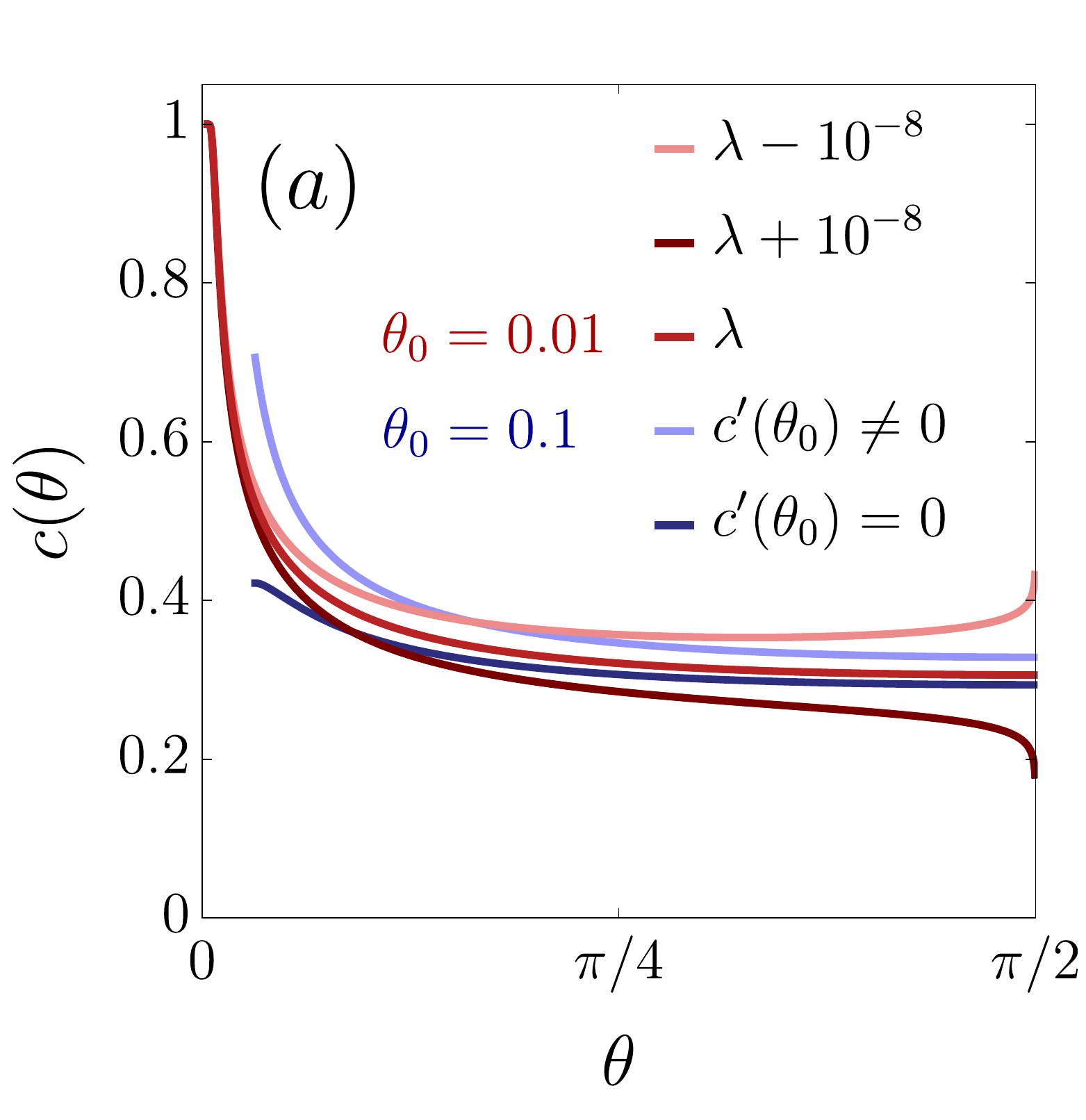}
    \includegraphics[width=.33\columnwidth]{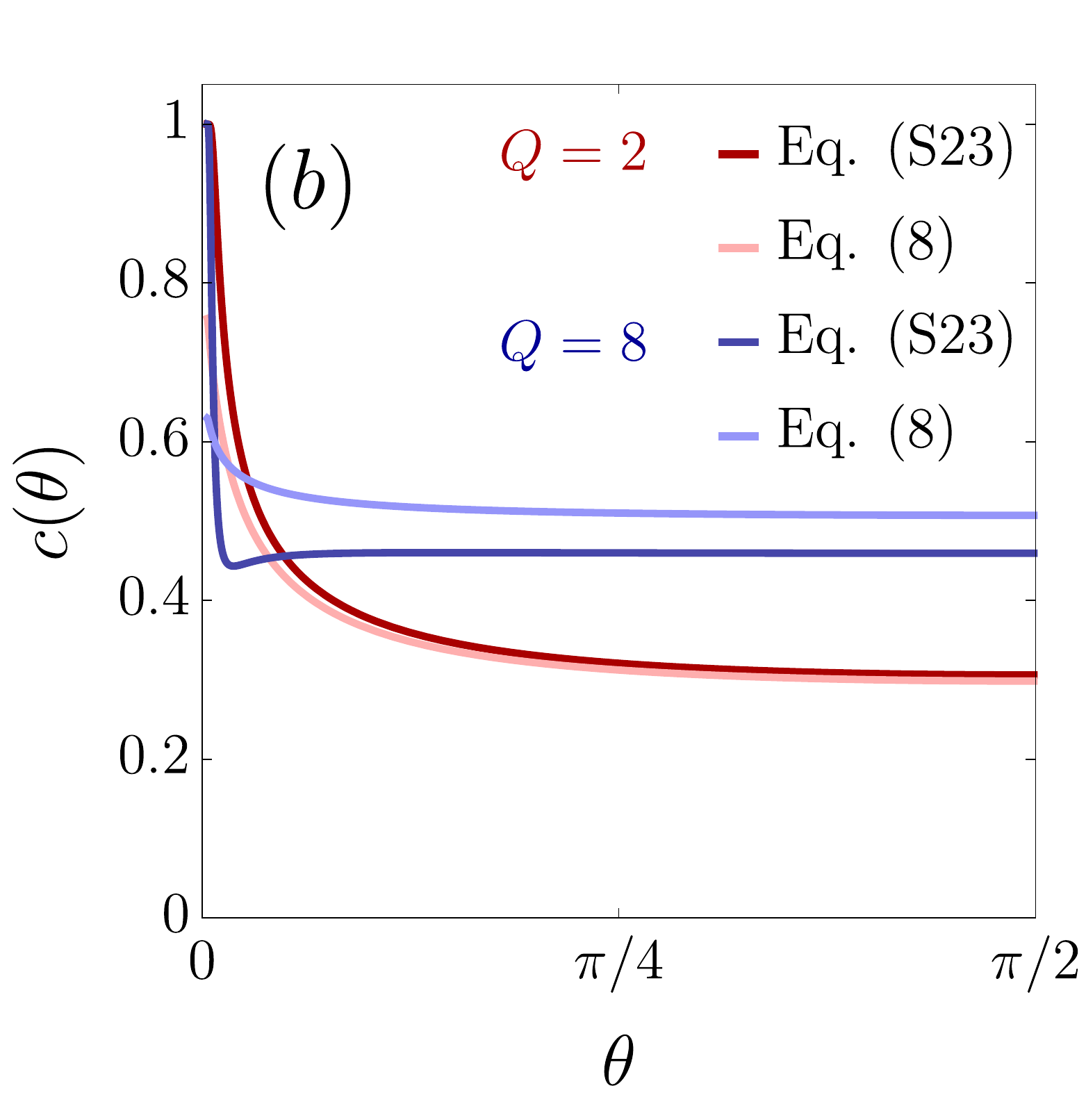}
    \caption{(a) Iterative solution of the differential equation \eqref{eq:difffull}. Parameters are 
    $\alpha=0.05$, $Q=2$, $\theta_0=0.2\alpha$. The converged solution is obtained for $\lambda=0.0098715246968$, and the solution is very sensitive to the initial conditions, deviating substantially already when changing $\lambda$ by only $10^{-8}$. In blue we show less precise solutions which have been obtained for a much larger $\theta_0$, for two different choices for $c'(\theta_0)$. Nevertheless, their deviation from the correct value at $\theta=\pi/2$ remains small.
    (b) Comparison between the exact differential equation \eqref{eq:difffull}, and the expansion for small $\alpha$ (Eq.~(8) of the main text), for a comparatively sizable $\alpha=0.05$. Both compare favorably for moderate values of $Q$. Only if  $Q\gg 1$, the solutions deviate noticeably. In all examples the specularity parameter is $q=1$.}
    \label{fig:fitting1}
\end{figure}

Unfortunately, this procedure still encounters issues at small $\alpha$ as the distribution function becomes exponentially close to $1$, which means that the initial condition $c(\theta_0)\approx 1$ is rendered increasingly sensitive to machine-precision limitations. We found two avenues to remedy this issue. 

One approach amounts to choosing a much larger $\theta_0$, which obviously cannot capture the form of $c(\theta)$ for angles below $\theta_0$, but converges smoothly to the correct solution at large $\theta$.
The second option is to expand Supplementary Eq.~\eqref{eq:difffull} in the limit $\alpha\to 0$. Doing so removes the exponential functions, and thus the essential singularities at $\theta=0$, so that the choice of very small $\theta_0$ remains unproblematic even for small $\alpha$.
Supplementary Fig.~\ref{fig:fitting1} shows the differences and commonalities between both approximations.

We calculate the limiting value $c(\pi/2)\equiv c_1$ in two steps. 
Using the expanded differential equation, Eq.~(8) in the main text, the canonical expansion of $c(\theta)$ around $\pi/2$ yields
\begin{align}
    c(\tfrac{\pi}{2})&=
    \frac{2 Q }{q}c''(\tfrac{\pi}{2})
    +
    \frac{2 - q + 2 Q }{q}\alpha
    \\
    c''(\tfrac{\pi}{2})&=
\frac{4 Q  }{3 q+20 Q}c^{(4)}(\tfrac{\pi}{2})
+\frac{  3 (4-q)}{3 q+20 Q}\alpha
    \\
    c^{(4)}(\tfrac{\pi}{2})&=
    \frac{2  Q (3 q+20 Q)}{5 q^2+120 q Q+488 Q^2}c^{(6)}(\tfrac{\pi}{2})
    +\frac{q (-25 q-234 Q+160)+1336 Q}{5 q^2+120 q Q+488 Q^2}\alpha
\end{align}
with all odd derivatives being equal zero. The expansion indicates that no non-analyticities are acquired at any finite order. On the other hand, as is visible in Supplementary Fig.~\ref{fig:fitting2}, $c_1/\alpha$ as a function of $\alpha$ reveals clearly contains a logarithmic divergence $c_1\propto \alpha\log\alpha^{-1}$. 
Based on the structure displayed in the sequence of  derivatives, we therefore make the ansatz that $c''(\pi/2)\approx \alpha\log(\lambda_1/\alpha)/(\lambda_2 q + \lambda_3 Q)$ with three fitting parameters $\lambda_1$, $\lambda_2$ and $\lambda_3$.

\begin{figure}
    \centering
    \includegraphics[width=.25\columnwidth]{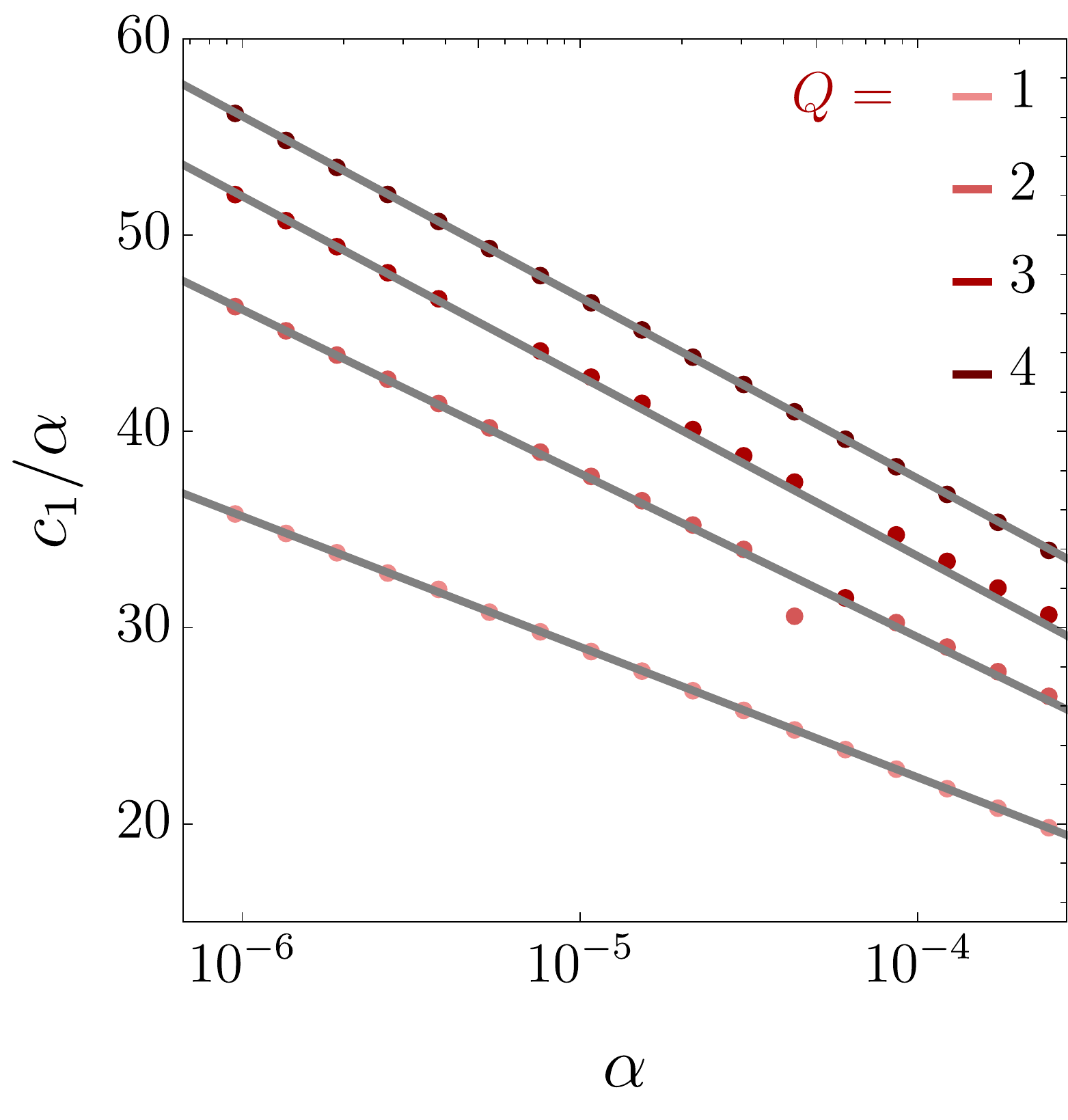}
    \hfill
    \includegraphics[width=.33\columnwidth]{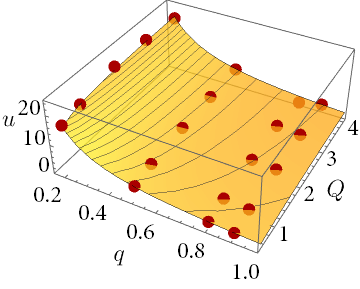}
    \hfill
    \includegraphics[width=.33\columnwidth]{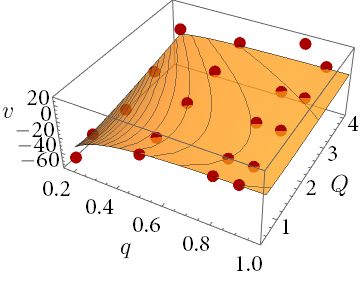}
    \caption{
    (Left) Logarithmic divergence of $c_1/\alpha$ in the limit $\alpha\to 0$, obtained from solving Eq.~(8) of the main text, for the parameter $q=0.9$. The gray curves show the fit with a logarithm, $c_1(\alpha)/\alpha=u\log\alpha^{-1} +v$.
    (Middle, Right) Results from fitting the relation $c_1(\alpha)/\alpha$ with the parameters $u$ ($v$) are shown in the middle (right). The red points are values of $u$ and $v$ as determined from numerically solving Eq.~(8) as a function of $\alpha$ for various values of $(q,Q)$. Overlaid are the functions Eqs.~(\ref{eq:u},\ref{eq:v}), for the parameter values $\lambda_2=0.39$,  $\lambda_3=0.46$ and $\lambda_1=0.061$.}
    \label{fig:fitting2}
\end{figure}

Then, we perform a numerical fit to $c_1(\alpha)/\alpha=u\log\alpha^{-1} +v$ for a range of values of $(q,Q)$. 
The result of this fit is shown in Supplementary Fig.~\ref{fig:fitting2}.
Obviously, the fitting parameters $u$ and $v$ are related to $\lambda_i$ via
\begin{align}
    u&=
    \frac{2Q}{q}\frac{1}{\lambda_2 q + \lambda_3 Q}
    \label{eq:u}
    \\
    v&=
    \frac{2 Q \frac{\log(\lambda_1)}{\lambda_2 q + \lambda_3 Q}+ (2 - q + 2 Q)}{q}
    \label{eq:v}
\end{align}
These relations are used to fit $\lambda_2=0.39$ and $\lambda_3=0.46$ from $u$ in Supplementary Eq.~\eqref{eq:u} and after fixing them, to also fit $\lambda_1=0.061$ from $v$, Supplementary Eq.~\eqref{eq:v}, thus yielding the approximate form of $c''(\pi/2)$ stated in the main text.

\end{document}